\documentclass[12pt,draftclsnofoot,onecolumn,doublespacing]{IEEEtran}
\usepackage{bbm}
\usepackage{amsfonts} 
\usepackage{mathrsfs}
\usepackage[cmex10]{amsmath}
\usepackage{amssymb}
\usepackage{graphicx}
\newtheorem{theorem}{Theorem}
\newtheorem{lemma}{Lemma}

\newcommand{\aeq}{\buildrel \textstyle (a)\over =}
\newcommand{\beq}{\buildrel \textstyle (b)\over =}

\newcommand{\cleq}{\buildrel \textstyle (c)\over \leq}

\begin{document}

\title{Cognitive Beamforming for Multiple Secondary Data Streams With Individual SNR Constraints}
\author{Sheng-Ming~Cai and Yi~Gong,~\IEEEmembership{Senior Member,~IEEE}

\thanks{S.-M. Cai is with the School
of Electrical and Electronic Engineering, Nanyang Technological
University, Singapore (e-mail: cais0010@ntu.edu.sg).}
\thanks{Y. Gong was with the School
of Electrical and Electronic Engineering, Nanyang Technological
University, Singapore. He is now with the South University of Science and Technology of China, Shenzhen, China (e-mail: jamesgong@ieee.org).}
}

\maketitle

\begin{abstract}
In this paper, we consider cognitive beamforming for multiple secondary data streams subject to individual signal-to-noise ratio (SNR) requirements
for each secondary data stream. In such a cognitive radio system, the secondary user is permitted to
use the spectrum allocated to the primary user as long as the caused interference at the primary receiver is tolerable. With both secondary SNR constraint and primary interference power constraint, we aim to minimize the secondary transmit power consumption.
By exploiting the individual SNR requirements, we
formulate this cognitive beamforming problem as an optimization problem on the Stiefel manifold. Both zero forcing beamforming (ZFB) and nonzero forcing beamforming (NFB) are considered. For the ZFB case,
we derive a closed form beamforming solution. For the NFB
case, we prove that the strong duality holds for the nonconvex primal problem and thus the optimal solution can be easily obtained by solving the dual problem. Finally, numerical results are presented to illustrate the performance of the proposed cognitive beamforming solutions.
\end{abstract}


\section{Introduction}
Cognitive radio (CR) has received considerable attention over the past few years because of its potential to ease the current overcrowded frequency spectrum. Based on the current spectrum allocation
policy, most frequency bands are allocated exclusively to specified services. However, such
policy results in underutilization of precious spectrum resources \cite{REF_FCC}. In the meantime, the demand of extra spectrum is increasing with the rapid growth of wireless applications. As a result, it is worth considering the idea of allowing other users to use the spectrum while guaranteeing the priority of authorized users. In a CR network, the spectrum can be shared with unauthorized or \emph{secondary} users (SUs), provided that they do not cause harmful interference to the authorized or \emph{primary} users (PUs) \cite{REF_HAYKIN,Lu}. SUs may transmit when they detect a spectrum hole in either time or frequency domain \cite{REF_SENSING1, REF_SENSING2}. Such schemes usually work when the spectrum is severely underutilized, otherwise SUs might not have sufficient opportunities to gain channel access. Therefore, the secondary throughput would be significantly constrained and the secondary system would suffer from a long latency.

Multiple-input multiple-output (MIMO) technology provides extra spatial dimensions
for transmissions. Multiple antennas
can be used to reduce the interference at PU
and satisfy the demand for high data rate at SU
by carefully designing transmit and receive beamforming \cite{REF_ISLAM_MIMO_CR}. As a result, SUs
may access the primary spectrum without causing harmful interference, even if PUs are also
using the spectrum at the same time. By assuming full channel knowledge known at SU, the capacity of a CR network is given in closed form in \cite{REF_RZHANG_PSVD}, when no interference is allowed at PU. For the case of nonzero interference power constraint, the expressions of the secondary signal-to-noise ratio (SNR) and the interference power received at PU usually
result in quadratically constrained quadratic
programming problems and these problems may not be directly solved by convex tools, especially when there is a rank constraint. Semidefinite programming (SDP) relaxation can be used to convert such problem to a convex optimization problem by dropping the rank constraint and
generate a local optimum \cite{REF_SDP}.  It is shown in \cite{REF_ZHANGYJ_SDR} that under certain conditions, a new solution can be generated from the one obtained by SDP relaxation without ruining the
constraints or changing the objective function, and hence the solution is optimal. However,  in general scenarios, the obtained local optimum
may not be feasible for the original problem because usually its rank
does not meet the requirement. As a result, approximation approaches such
as the randomization procedure are used to generate
a feasible solution \cite{REF_LUOZQ_SDR}, \cite{REF_SDR_APP1}.

Under individual SNR constraints, the downlink transmission where each user has a single data stream is studied in \cite{REF_SCHUBERT_INDIV_SINR}. In \cite{REF_KHACHAN_INDIV_SINR}, the study is extended to multiple data streams. However, there exists interference between any two data streams even if they are for the same user. In \cite{REF_YHYANG_INDIV_SINR}, the authors studied the transmit power minimization problem with individual SNR requirements and used joint decoding to remove the interference. An iterative algorithm is proposed therein to solve the problem, but it is not clear whether the iterative algorithm can converge to the global/local optimum. The cognitive transmission with multiple antennas equipped at secondary transmitter (ST) and secondary receiver (SR) is studied in \cite{REF_HAOY_STM}, where the secondary throughput is maximized subject to the secondary power constraint and primary interference power constraint. It is shown therein that the secondary transmit beamforming problem can be converted to an optimization problem with unitary constraint, and then an algorithm is proposed to compute the beamforming matrix such that a local optimum can be obtained.

In this paper, we study the problem of secondary transmit beamforming with multiple secondary data streams subject to individual SNR constraints. We are interested in the beamforming design at ST so as to minimize its transmit power
under both its own per data stream SNR constraint at SR and the interference-power constraint at the primary receiver (PR). We use zero forcing beamforming (ZFB) to deal with the case when no interference is allowed at PR. If a positive interference power constraint is allowed, we design nonzero forcing beamforming (NFB). We formulate the secondary power minimization problem as an optimization problem on the Stiefel manifold \cite{REF_MANTON}. We show that SDP relaxation can achieve the  global optimum when there is a
single data stream but it may not be suitable for the multiple data streams scenario.
For the multiple data streams, we derive a closed form solution for the ZFB case.
As for the NFB case, we analyze the associated dual problem and provide the sufficient
condition for strong duality. As a result, the global optimum can
be obtained by solving the dual problem efficiently.

The rest of this paper is organized as follows. In Section II, we describe the system model and formulate the main optimization problem. The secondary beamforming feasibility test is also provided in this section.
In Section III, we prove that with a single secondary data stream, SDP relaxation can
lead to the global optimum. The case of multiple secondary data streams is considered in Section IV, where we firstly derive the closed-form solution for the ZFB problem. We then prove that for the NFB problem, the strong duality holds as well as the optimality of the solution of the dual problem.
The numerical results are provided in Section V to illustrate the performance of the proposed secondary beamforming solutions. This paper is concluded in Section VI.

\emph{Notations:} Scalar is denoted by a lower-case letter, while
vector is denoted by bold-face lower-case letter and matrix is denoted
by bold-face upper-case letter. $\textbf{I}_p$ denotes the
$p \times p$ identity matrix. For a matrix \(\textbf{S}\),
tr\((\textbf{S})\), rank\((\textbf{S})\), \(\textbf{S}^H\), and $S_{ij}$ denote its
trace, rank, Hermitian matrix, and the entry at the $i$-th row and the
$j$-the column, respectively.
\(\mathrm{diag}(s_1, s_2, \cdots, s_n)\) denotes a diagonal matrix
with diagonal elements given by \(s_1, s_2, \cdots, s_n\). For a matrix $\textbf{S}$, $\textbf{S}\succeq \textbf{0}$ denotes that $\textbf{S}$ is positive semidefinite.
The complex Stiefel manifold \(St(n,p)\) is
the set \(St(n,p) = \{\textbf{V} \in \mathbb{C}^{n \times p}:
\textbf{V}^H \textbf{V} = \textbf{I}_p\}\), where \(n \geq p\).
For a Hermitian matrix \(\textbf{S} \in \mathbb{C}^{n \times n}\),
the eigenvalue decomposition (EVD) is represented as \(\textbf{S} =
\textbf{U}^H \mathbf{\Sigma} \textbf{U}\), where $\textbf{U}$ is a
unitary matrix and $\mathbf{\Sigma}$ is a diagonal matrix. For a
Hermitian matrix \(\textbf{S} \in \mathbb{C}^{n \times n}\) with
$\mathrm{rank}(\textbf{S}) = r \leq n$, \(\textbf{S}^{-1}\) denotes the
Moore-Penrose pseudoinverse of $\textbf{S}$, i.e., $\textbf{S} \textbf{S}^{-1} \textbf{S} = \textbf{S}$
and $\textbf{S}^{-1} \textbf{S} \textbf{S}^{-1} = \textbf{S}^{-1}$. It is obtained
from the following decomposition of $\textbf{S}$: If \(\textbf{S}\) is written as
\(\textbf{S} = \textbf{U}^H \mathbf{\Sigma} \textbf{U}\), where
\(\textbf{U}^H \in St(n,r)\), \(\mathbf{\Sigma} \in \mathbb{R}^{r \times r}\)
is a full rank diagonal matrix, then $\textbf{S}^{-1} = \textbf{U}^H
\mathbf{\Sigma}^{-1} \textbf{U}$. Likewise, $\textbf{S}^{1/2}$ and
$\textbf{S}^{-1/2}$ are found as
\(\textbf{S}^{1/2} = \textbf{U}^H \mathbf{\Sigma}^{1/2} \textbf{U}\) and
\(\textbf{S}^{-1/2} = \textbf{U}^H \mathbf{\Sigma}^{-1/2} \textbf{U}\), respectively.

\section{System Model and Problem Formulation}

\subsection{System Model}

We consider a multi-antenna CR network in the presence of primary transmission, where there are a single pair of ST and SR, supporting multiple secondary data streams, and the primary users and the secondary users share the same bandwidth for transmission in an overlay approach. In particular, both the interference caused by secondary transmission experienced at PR and the interference caused by primary transmission experienced at SR are considered in this paper. We consider narrowband transmission for both primary and secondary users where multiple antennas are equipped at the primary transmitter (PT), PR, ST and SR. The numbers of antennas equipped at PT, PR, ST and SR
are denoted as \(p\), \(q\), $m$ and \(n\), respectively. Let $\textbf{G}_\textbf{x} \in \mathbb{C}^{n \times p}$, $\textbf{H}_\textbf{x} \in \mathbb{C}^{q \times m}$ and $\textbf{H} \in \mathbb{C}^{n \times m}$ denote the channel matrices (all assumed to be full rank) of the PT \(\rightarrow\) SR link, ST \(\rightarrow\)  PR
link and ST \(\rightarrow\) SR link, respectively. It is assumed that $\textbf{H} \in \mathbb{C}^{n \times m}$ and $\textbf{H}_\textbf{x} \in \mathbb{C}^{q \times m}$ are known at ST, and $\textbf{G}_\textbf{x} \in \mathbb{C}^{n \times p}$ is known at SR. Under this assumption, subject to its own SNR constraint at each data stream, ST is able to adjust its beamforming matrix based on the channel knowledge so as to optimally balance between minimizing its own transmit power and avoiding interferences
at PR. In a fading environment, there are cases where it
is difficult for ST and SR to obtain perfect knowledge of the
instantaneous channels. In such cases, the results obtained in
this paper provide a performance upper-bound for the considered secondary
transmit beamforming problem.

Letting \(\textbf{s}_{p} \in \mathbb{C}^{p \times 1}\) denote the transmitted
primary signal with zero mean and variance $P_{p}$ and
\(\textbf{s}_{s} \in \mathbb{C}^{m \times 1}\) denote the transmitted secondary signal,
 the received signal at
SR can be written as
\setlength\arraycolsep{1.4pt}
\begin{equation}
\textbf{y} = \textbf{H} \textbf{s}_{s} + \textbf{G}_\text{x} \textbf{s}_{p} + \textbf{z}
\end{equation}
where \(\textbf{z} \in \mathbb{C}^{n \times 1}\) represent the additive Gaussian
noise with zero mean and unit variance at SR. 
The second term on the right-hand side of (1) represents the interference
from the primary transmission. Therefore, the interference-plus-noise
covariance matrix at SR is given by
\begin{equation}
\textbf{W} =
\text{E}[(\textbf{G}_\text{x} \textbf{s}_{p} + \textbf{z})
(\textbf{G}_\text{x} \textbf{s}_{p} + \textbf{z})^H]
= {P}_{p} \textbf{G}_\text{x} \textbf{G}^H_\text{x} + \textbf{I}_n . \nonumber
\end{equation}

Letting $\textbf{T} \in \mathbb{C}^{m \times d}$ denote the secondary transmit
beamforming matrix,
the transmitted secondary signal can be represented as
\begin{equation}
\textbf{s}_{s} = \textbf{T} \textbf{d}_{s} \nonumber
\end{equation}
where $\textbf{d}_{s} \in \mathbb{C}^{d \times 1}$ denotes the
secondary data, modeled as a random vector with
$d \leq \min(m,n)$ denoting the number of secondary data streams and $\mathrm{E}[\textbf{d}_{s}
\textbf{d}_{s}^H] = \textbf{I}_d$.

It can be easily shown that the eigenvalues of $\textbf{T}^H
\textbf{H}^H \textbf{W}^{-1} \textbf{H} \textbf{T}$ represent the SNR\footnote{In this paper, it actually refers to the signal-to-interference-plus-noise ratio.} values
of secondary data streams at SR, after proper receive beamforming that maximizes the SNR of each data stream, as shown in Appendix A.
Letting $\textbf{M} \triangleq \textbf{H}^H \textbf{W}^{-1} \textbf{H}$, we can use
EVD to decompose $\textbf{T}^H \textbf{M} \textbf{T}$ as
$\textbf{T}^H \textbf{M} \textbf{T} =
\textbf{U}^H \mathbf{\Sigma} \textbf{U}$,
where the diagonal entries of
\(\mathbf{\Sigma}\) now represent the SNR of each secondary data stream.
We can always choose a unitary matrix and post-multiply it to $\textbf{T}$ to get a new $\textbf{T}$ such
that $\textbf{T}^H \textbf{M} \textbf{T}$ is diagonal with the same eigenvalues, i.e.
\begin{equation}
\textbf{T}^H \textbf{M} \textbf{T} = \mathbf{\Sigma}.
\label{eq3}
\end{equation}

In order to protect the primary communication, the interference power experienced at
PR should not exceed a certain threshold. The peak interference power
constraint can then be written as
\begin{equation}
\text{tr}(\textbf{T}^H \textbf{H}_\text{x}^H \textbf{H}_\text{x}
\textbf{T}) \leq \xi  \nonumber
\end{equation}
where the value of $\xi$ represents
the maximum tolerable interference power at PR.
As $\xi$ increases, ST has
higher flexibility to design the transmit beamforming matrix. If $\xi$ is sufficiently
large, ST can communicate to SR as if PR is absent. For a certain $\xi$, it is possible that the underlying channel conditions fail to support the secondary QoS requirement with a certain $d$. In this case, ST may have to reduce $d$ or relax its QoS requirement to be able to transmit. Given the secondary QoS requirement and primary interference power constraint, we can test if there is a feasible secondary beamforming solution.

Clearly, the number of secondary data
streams $d$ should not be greater than $\min(m,n)$. If the number of ST antennas, \(m\), is strictly larger than the
number of PR antennas, \(q\), then there are
\(m-q\) available degrees of freedom or spatial
dimensions for secondary transmission without causing any interference at PR, which
can be realized by placing $\textbf{T}$ in the null space of
$\textbf{H}_\text{x}$. On the other hand, if the PR can tolerate a nonzero interference (i.e., $\xi > 0$), the number of supported secondary data streams
can be greater than $m-q$, depending on the value of $\xi$ and the
underlying channel condition. Therefore, the secondary system can support at
least $m-q$ secondary data streams.
In this paper, we consider individual SNR requirements for all the
secondary data streams. Let $\rho_i$ denote the $i$-th data stream's SNR requirement, where $i = 1, \cdots, d$. Without loss of generality,
we assume $\rho_1 \geq \cdots \geq \rho_d$.

\subsection{Problem Formulation}
Our objective is to minimize the secondary sum transmit power while satisfying both the secondary per data stream
SNR constraint and the primary interference power constraint. Such
problem is formulated as
\begin{subequations}
\begin{eqnarray}
\label{OBJ_P1}
\!\!\!\!\!\!\!\!\!\!\!\!\!\!\!\!\!\!\!\!\!\!\!\!\!\!\!(\text{P1}):\ \ \ \ \ \ \ \ \ \min_{\mathbf{T}} && \text{tr}(\textbf{T}^H \textbf{T})\\
\label{CONST1_P1}
\mathrm{s. t.} & &\text{tr}(\textbf{T}^H \textbf{H}_\text{x}^H
\textbf{H}_\text{x} \textbf{T}) \leq \xi \\
\label{ACHIV_SNR}
&& \mathbf{\Sigma} \succeq \mathrm{diag}(\rho_1, \rho_2, \cdots, \rho_{d}).
\end{eqnarray}
\end{subequations}

\begin{lemma}
\label{Lemma_1}
\emph{Any} $\textbf{T}$ \emph{satisfying
(\ref{eq3}) can be expressed as}
\begin{equation}
\textbf{T} = \textbf{M}^{-1/2} \textbf{V} \mathbf{\Sigma}^{1/2}
\label{FORM_T_S}
\end{equation}
\emph{where} $\textbf{V} \in St(m,d)$.
\end{lemma}

\begin{IEEEproof}
Please refer to Appendix B.
\end{IEEEproof}

\begin{lemma}
The inequality constraint in \eqref{ACHIV_SNR} can be replaced with its equality constraint, i.e., $\mathbf{\Sigma} = \mathrm{diag}(\rho_1, \cdots, \rho_{d})$.
\end{lemma}

\begin{IEEEproof}
With (\ref{FORM_T_S}), the objective function in (\ref{OBJ_P1}) and the constraint in (\ref{CONST1_P1}) can be respectively rewritten as
\begin{eqnarray}
\text{tr}(\textbf{T}^H \textbf{T}) &=&
\text{tr}(({\mathbf{\Sigma}^{1/2}})^H \textbf{V}^H
(\textbf{M}^{-1/2})^H \textbf{M}^{-1/2} \textbf{V}
{\mathbf{\Sigma}^{1/2}})\nonumber \\
&=& \text{tr}({\mathbf{\Sigma}} \textbf{V}^H
\textbf{M}^{-1} \textbf{V})
\label{NEW_POWER}
\end{eqnarray}
and
\begin{equation}
\text{tr}(\textbf{T}^H \textbf{H}_\text{x}^H \textbf{H}_\text{x}
\textbf{T})= \text{tr}({\mathbf{\Sigma}} \textbf{V}^H \textbf{M}_\text{x}
\textbf{V}) \leq \xi
\label{NEW_INT}
\end{equation}
where $\textbf{M}_\text{x} \triangleq \textbf{M}^{-1/2} \textbf{H}_\text{x}^H
\textbf{H}_\text{x} \textbf{M}^{-1/2}$.

Since it always holds that
\begin{eqnarray}
\text{tr}(\mathbf{\Sigma} \textbf{V}^H
\textbf{M}^{-1} \textbf{V}) &\geq& \text{tr}({\mathrm{diag}(\rho_1, \cdots, \rho_d)} \textbf{V}^H
\textbf{M}^{-1} \textbf{V}) \nonumber \\
\text{tr}(\mathbf{\Sigma} \textbf{V}^H \textbf{M}_\text{x}
\textbf{V}) &\geq& \text{tr}({\mathrm{diag}(\rho_1, \cdots, \rho_d)} \textbf{V}^H \textbf{M}_\text{x}
\textbf{V}) \nonumber
\end{eqnarray}
we can set $\mathbf{\Sigma} = \mathrm{diag}(\rho_1,\cdots,\rho_d)$ without affecting the optimal solution of (P1).
\end{IEEEproof}

With \eqref{NEW_POWER}, \eqref{NEW_INT}, and Lemma 2, the original problem (P1) can be reformulated as
\begin{subequations}
\begin{eqnarray}
\label{OBJ_P2}
\!\!\!\!\!\!\!\!\!\!\!\!\!\!\!\!\!\!\!\!\!\!\!\!\!\!\!\!\!\!\!(\text{P2}):\ \ \ \ \ \ \ \ \min_{\textbf{V} \in St(m,d)} && \text{tr}(\mathbf{\Sigma} \textbf{V}^H \textbf{M}^{-1}
\textbf{V}) \\
\label{CONST1_P2}
\mathrm{s. t.} && \text{tr}(\mathbf{\Sigma} \textbf{V}^H \textbf{M}_\text{x}
\textbf{V}) \leq \xi.
\end{eqnarray}
\end{subequations}

\subsection{Access Feasibility}
Before solving (P2), we need to perform a feasibility test. Note that by letting $\mathbf{\Sigma} = \mathrm{diag}(\rho_1,\cdots,\rho_d)$ and expressing $\textbf{T}$ in the form of \eqref{FORM_T_S}, we have already met the secondary per data stream SNR requirement. However, for a given primary interference power constraint ($\xi$) and underlying channel condition ($\textbf{M}_\text{x}$), \eqref{CONST1_P2} is not always satisfied. If no $\textbf{V} \in St(m,d)$ satisfies the interference constraint \eqref{CONST1_P2}, we say that the secondary transmission is not \emph{feasible}.

We define
\begin{equation}
\xi_0 \triangleq \min_{\textbf{V} \in St(m,d)} \text{tr}(\mathbf{\Sigma} \textbf{V}^H
\textbf{M}_\text{x} \textbf{V}).
\end{equation}
If the actual interference power constraint $\xi \geq \xi_0$, we can
find a feasible $\textbf{V}$ to generate the secondary transmit beamforming matrix
to satisfy both the secondary SNR and primary interference power constraint, otherwise ST should keep silent.

In order to find $\xi_0$, we provide the following lemma first.
\begin{lemma}
\label{THEOREM_MIN_TRACE_PRODUCT}
\emph{Given a diagonal matrix $\mathbf{\Delta} = \mathrm{diag}(\delta_1, \cdots, \delta_u)$
$\in \mathbb{C}^{u \times u}$ $(\delta_1 \geq \cdots
\geq \delta_u)$ and a Hermitian matrix $\mathbf{\Omega} \in \mathbb{C}^{v \times v}$ $(v \geq u)$ with  $\omega_i$, $i = 1, \cdots, v$, being its the eigenvalues $(\omega_1 \geq \cdots \geq \omega_v)$ and $\boldsymbol{\varphi}_i$ being its eigenvector corresponding to $\omega_i$, for a matrix $\mathbf{\Theta} \in St(v,u)$, we have the following inequality:}
\begin{equation}
\text{tr} (\mathbf{\Delta} \mathbf{\Theta}^H \mathbf{\Omega} \mathbf{\Theta})\geq \sum_{i = 1}^u \delta_i \omega_{v-i+1}
\end{equation}
\end{lemma}
where the equality holds if $\mathbf{\Theta}$ is constructed as $\mathbf{\Theta} = [\boldsymbol{\varphi}_v,
\cdots,  \boldsymbol{\varphi}_{v-u+1}]$.
\begin{IEEEproof}
Please refer to Appendix C.
\end{IEEEproof}

With Lemma \ref{THEOREM_MIN_TRACE_PRODUCT}, we can show that $\xi_0$ is given by
\begin{equation}
\label{ACHIV_MIN_INTF}
\xi_0 = \sum_{i = 1}^d \rho_i x_{m-i+1}
\end{equation}
where $x_i$ denotes the $i$-th eigenvalue of $\textbf{M}_\text{x}$ and
$x_1 \geq \cdots \geq x_m$. Since $\text{rank}(\textbf{M}_\text{x}) = q$, we have
\begin{eqnarray}
\label{CROSSCHANNEL_EIG}
\left\{
\begin{array}{cc}
x_i > 0, & i \leq q \\
x_i = 0, & i > q. \\
\end{array}
\right.
\end{eqnarray}

It is clear from \eqref{ACHIV_MIN_INTF} and \eqref{CROSSCHANNEL_EIG} that when $m-q \geq d$, $\xi_0=0$ and therefore secondary access (via ZFB or NFB) is always possible; when $m-q < d$, $\xi_0$ will always be greater than zero and therefore only NFB is possible.

\begin{figure}[!t]
\centering
  \includegraphics[width=4.9in]{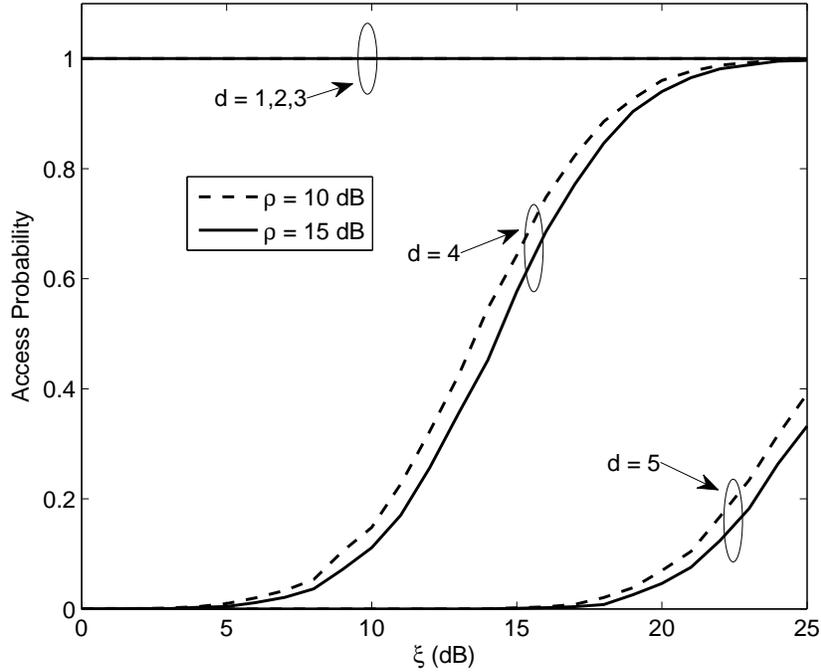}
  \caption{Secondary access probability, i.e., the probability of $\xi_0 \leq \xi$. $m=n=5$, $p=q=2$ and $\rho=\rho_1=...=\rho_d$.}
  \label{AccessProb}
\end{figure}

In Fig. \ref{AccessProb}, we illustrate the secondary access probability, i.e., the probability of $\xi_0 \leq \xi$, with different number of secondary data streams, where identical SNR requirements are considered for multiple-stream cases. It is assumed that each entry of the involved channel matrices is an i.i.d. random variable, distributed as $\mathcal{CN}(0,1)$. With $m=n=5$ and $p=q=2$, Fig. \ref{AccessProb} shows that the cases of $d=1$, 2 and 3 lead to $100\%$ access probability since $m-q \geq d$ in these cases. On the other hand, when $d>3$, the access probability, increasing with a lower secondary SNR requirement, heavily depends on the number of secondary streams and  the primary interference constraint $\xi$. It is obvious that in this case more data streams or lower primary interference constraint results in a lower secondary access probability.

\section{The Case of Single Data Stream}

In this section, we show that SDP relaxation can be used to find the optimal secondary beamforming solution for the single data stream case.

By defining $\textbf{X} \triangleq \textbf{T} \textbf{T}^H$ and
dropping the rank constraint (rank($\textbf{X}$) = rank($\textbf{T}$) = $d$),
(P1) can be reformulated as a relaxed problem
\begin{subequations}
\begin{eqnarray}
\!\!\!\!\!\!\!\!\!\!\!\!\!(\text{P3}):\ \ \ \ \min_{\textbf{X}\succeq \textbf{0}} && \text{tr}(\textbf{X})\\
\label{CONST1_SDR}
\mathrm{s. t.} && \text{tr}(\textbf{H}_{\text{x}}^{H}
\textbf{H}_{\text{x}} \textbf{X}) \leq \xi \\
\label{CONST2_SDR}
&& \text{det}(\textbf{X} \textbf{M} -
\rho_i \textbf{I}_n) = 0, \forall i = 1, \cdots, d.
\end{eqnarray}
\end{subequations}

The secondary per data SNR requirement is reflected in \eqref{CONST2_SDR}, since
$\textbf{X} \textbf{M}$ shares the
same eigenvalues with $\textbf{T}^H \textbf{M} \textbf{T}$ and
these eigenvalues denote the required secondary SNR values.

When there is a single data stream, constraint \eqref{CONST2_SDR} can be equivalently rewritten
as $\text{tr}(\textbf{X} \textbf{M}) = \rho_1$. It turns out that all the constraints together with the objective function
are convex and thus we can apply any convex optimization algorithm to solve the problem.
Let $\textbf{X}^\ast$ denote the solution for (P3).
Since the problem is relaxed, $\textbf{X}^\ast$ leads to the optimal beamforming vector
if $\text{rank}(\textbf{X}^\ast) = 1$, or another rank-one Hermitian matrix can be generated
from $\textbf{X}^\ast$ with all the optimization objective function and constraints unchanged \cite{REF_LUOZQ_SDR}.
Otherwise, there will be a nonzero gap between the solutions of the relaxed problem (P3) and the
original problem (P1).

\begin{theorem}
\label{THEOREM_SDP_RANKONE_OPT}
\emph{When $d = 1$, there exists a rank-one solution of the relaxed problem that is optimal.}
\end{theorem}

\begin{IEEEproof}
The dual of the relaxed single data stream problem is given by
\begin{subequations}
\begin{eqnarray}
\label{OBJ_DUAL}
\min_{\mu_1, \mu_2 \in \mathbb{R}} && \mu_1 \xi + \mu_2 \rho_1 \\
\label{CONST1_DUAL}
\mathrm{s. t.} && (\textbf{I} + \mu_1 \textbf{H}_\text{x}^H \textbf{H}_\text{x} +
\mu_2 \textbf{M}) \succeq \textbf{0} \\
&& \mu_1 \geq 0
\end{eqnarray}
\end{subequations}
where $\mu_1$ and $\mu_2$ are the Lagrange multipliers associated with \eqref{CONST1_SDR} and
$\text{tr}(\textbf{X} \textbf{M}) = \rho_1$, respectively. Since the relaxed
problem is convex, strong duality holds, i.e., the original and dual problems
lead to the same solution. As a result, the following complementary conditions establish
\begin{subequations}
\label{WHOLE_COND_COMPLEMENTARY}
\begin{eqnarray}
\label{COND1_COMPLEMENTARY}
\text{tr}(\textbf{X}^\ast (\textbf{I} + \mu_1^\ast \textbf{H}_\text{x}^H \textbf{H}_\text{x} +
\mu_2^\ast \textbf{M})) &=& {0} \\
\label{COND2_COMPLEMENTARY}
\mu_1^\ast (\text{tr}(\textbf{X}^\ast \textbf{H}_\text{x}^H \textbf{H}_\text{x}) - \xi) &=& 0 \\
\label{COND3_COMPLEMENTARY}
\mu_2^\ast (\text{tr}(\textbf{X}^\ast \textbf{M}) - \rho_1) &=& 0
\end{eqnarray}
\end{subequations}
where $\textbf{X}^\ast$
and $(\mu_1^\ast, \mu_2^\ast)$ are the primal and dual optimal solutions, respectively \cite{REF_BOYD_CVXOPT}.

If $\text{rank}(\textbf{X}^\ast) = 1$, $\textbf{X}^\ast$ can be decomposed as $\textbf{X}^\ast
= {\textbf{t}} {\textbf{t}}^H$, and hence ${\textbf{t}}\in \mathbb{C}^{m \times 1}$
is the optimal transmit beamforming vector.

If $\text{rank}(\textbf{X}^\ast) > 1$, based on the rank-one decomposition theory \cite{REF_SZZHANG_RANKONEDECOMP}, we can find a vector $\textbf{t} \in \mathbb{C}^{m \times 1}$ such that the following equations
\begin{subequations}
\begin{eqnarray}
\label{eq_1a}
\text{tr} (\textbf{t} \textbf{t}^H) &=& \text{tr}(\textbf{X}^\ast) \\
\label{eq_1b}
\text{tr}(\textbf{t} \textbf{t}^H \textbf{H}_\text{x}^H \textbf{H}_\text{x}) &=& \text{tr}(\textbf{X}^\ast \textbf{H}_\text{x}^H
\textbf{H}_\text{x}) \\
\label{eq_1c}
\text{tr} (\textbf{t} \textbf{t}^H \textbf{M}) &=&
\text{tr}(\textbf{X}^\ast \textbf{M})
\end{eqnarray}
\end{subequations}
hold simultaneously [9, Section V.B]. 

It can be directly concluded from \eqref{eq_1b} and \eqref{eq_1c} that matrix $\textbf{t} \textbf{t}^H$ satisfies conditions \eqref{COND2_COMPLEMENTARY} and
\eqref{COND3_COMPLEMENTARY}. That is
\begin{eqnarray}
\mu_1^\ast (\text{tr} (\textbf{t} \textbf{t}^H \textbf{H}_\text{x}^H \textbf{H}_\text{x}) - \xi) &=& 0 \nonumber \\
\mu_2^\ast (\text{tr} (\textbf{t} \textbf{t}^H \textbf{M})  - \rho_1) &=& 0. \nonumber
\end{eqnarray}

On the other hand, since
\begin{eqnarray}
&& \text{tr} (\textbf{t} \textbf{t}^H (\textbf{I} + \mu_1^\ast \textbf{H}_\text{x}^H \textbf{H}_\text{x} +
\mu_2^\ast \textbf{M})) \nonumber \\
&=& \text{tr} (\textbf{t} \textbf{t}^H) + \mu_1^\ast \text{tr}(\textbf{t} \textbf{t}^H \textbf{H}_\text{x}^H \textbf{H}_\text{x}) + \mu_2^\ast \text{tr} (\textbf{t} \textbf{t}^H \textbf{M}) \nonumber \\
&=& \text{tr}(\textbf{X}^\ast) + \mu_1^\ast \text{tr}(\textbf{X}^\ast \textbf{H}_\text{x}^H \textbf{H}_\text{x}) +
\mu_2^\ast  \text{tr}(\textbf{X}^\ast \textbf{M}) \nonumber \\
&=& \text{tr}(\textbf{X}^\ast (\textbf{I} + \mu_1^\ast \textbf{H}_\text{x}^H \textbf{H}_\text{x} +
\mu_2^\ast \textbf{M})) \nonumber \\
&=& 0 \nonumber
\end{eqnarray}
it follows that matrix $\textbf{t} \textbf{t}^H$ satisfies condition \eqref{COND1_COMPLEMENTARY}. Therefore, $\textbf{t} \textbf{t}^H$ satisfies all the three complementary conditions in \eqref{WHOLE_COND_COMPLEMENTARY}, and thus is the optimal rank-one solution of (P3). As a result, $\textbf{t}$ is the optimal secondary transmit beamforming vector resulting in zero duality gap.
\end{IEEEproof}

For the multiple data streams case, since constraint \eqref{CONST2_SDR} is nonconvex, to the best of our knowledge, no applicable reformulation
or relaxation on this constraint set can be found. As a result, the SDP relaxation might not be feasible for the considered multiple data streams case. In the next section, we reformulate the cognitive beamforming problem for multiple secondary data streams to a new problem on the Stiefel manifold and solve it effectively.

\section{The Case of Multiple Data Streams}

\subsection{Zero Forcing Beamforming}
In the ZFB scenario, no interference is allowed at PR. According to \eqref{ACHIV_MIN_INTF} and
\eqref{CROSSCHANNEL_EIG}, this scenario is possible only when $m-q \geq d$. Therefore, the primary interference constraint \eqref{CONST1_P2} is rewritten as
\begin{equation}
\label{EQN_TEMP1}
\textbf{H}_\text{x} \textbf{T} = \textbf{0}.
\end{equation}

The ZFB constraint \eqref{EQN_TEMP1} requires that the transmit beamforming matrix $\textbf{T}$
should be projected to the null space of $\textbf{H}_\text{x}$. By substituting
(\ref{FORM_T_S}) and the singular value decomposition (SVD) of
\(\textbf{H}_\text{x} = \textbf{U}_1 \textbf{A}_1 \textbf{V}_1\) into (\ref{EQN_TEMP1}), where \(\textbf{U}_1\) is a
\(q \times q\) unitary matrix, \(\textbf{A}_1\) is a positive definite
diagonal matrix, and \(\textbf{V}_1^H \in St(m,q)\), we have
\begin{equation}
\label{TEMP_4}
\textbf{H}_\text{x} \textbf{T} = \textbf{U}_1 \textbf{A}_1 \textbf{V}_1 \textbf{M}^{-1/2} \textbf{V} \mathbf{\Sigma}^{1/2} = \textbf{0} .
\end{equation}
It is clear to see that matrices \(\textbf{U}_1\)
and \(\textbf{A}_1\) are full-rank square
matrices. We remove them by left multiplying the corresponding inverse matrices and then drop the
SNR requirement matrix $\mathbf{\Sigma}$. It thus follows that
\begin{equation}
\label{TEMP_17}
\textbf{V}_1 \textbf{M}^{-1/2} \textbf{V} = \textbf{0} .
\end{equation}

Now we denote
the SVD of \(\textbf{V}_1 \textbf{M}^{-1/2}\) as
\(\textbf{V}_1 \textbf{M}^{-1/2} = \textbf{U}_2 \textbf{A}_2 \textbf{V}_2\),
where \(\textbf{U}_2\) is a \(q \times q\) unitary matrix, \(\textbf{A}_2\)
is a positive definite diagonal matrix, and \(\textbf{V}_2^H \in St(m,q)\).
Similarly, (\ref{TEMP_17}) establishes when
\begin{equation}
\textbf{V}_2 \textbf{V} = \textbf{0}.
\label{TEMP_1}
\end{equation}

The \(m\)-dimensional space that $\textbf{V}$ lies in can be separated
into two subspaces via the projected-channel SVD \cite{REF_RZHANG_PSVD}: one is perpendicular to \(\textbf{V}_2\)
(by multiplying (\(\textbf{I}_m - \textbf{V}_2^H \textbf{V}_2\))) and the
other is parallel to \(\textbf{V}_2\) (by multiplying
\(\textbf{V}_2^H \textbf{V}_2\)). Because
\(\text{rank}(\textbf{V}_2^H \textbf{V}_2) = \text{rank} (\textbf{V}_2) = q\)
and $\textbf{V}_2 (\textbf{I}_m - \textbf{V}_2^H \textbf{V}_2) = \textbf{0}$,
we have $\text{rank}(\textbf{I}_m - \textbf{V}_2^H \textbf{V}_2) = m - q$.
From \eqref{TEMP_1}, we can apply the subspace separation on \(\textbf{V}\) to get
\setlength\arraycolsep{1.4pt}
\begin{eqnarray}
& \textbf{V} &= [(\textbf{I}_m - \textbf{V}_2^H \textbf{V}_2) + \textbf{V}_2^H \textbf{V}_2] \textbf{V} \nonumber \\
& &= (\textbf{I}_m - \textbf{V}_2^H \textbf{V}_2) \textbf{V}.
\end{eqnarray}
Therefore, we can rewrite the ZFB problem as
\begin{subequations}
\begin{eqnarray}
\label{OBJ_P3}
\!\!\!\!\!\!\!\!\!\!\!\!\!\!\!\!\!\!\!\!\!\!\!\!\!\!\!\!(\text{P4}):\ \ \ \ \ \ \ \min_{\textbf{V} \in St(m,d)}&&\text{tr}(\mathbf{\Sigma} \textbf{V}^H
\textbf{M}^{-1} \textbf{V}) \\
\label{CONST_V_s}
\mathrm{s. t.} &&\textbf{V} = (\textbf{I}_m - \textbf{V}_2^H
\textbf{V}_2) \textbf{V}.
\end{eqnarray}
\end{subequations}

The problem is now reduced to finding a Stiefel manifold matrix
$\textbf{V}$ with a signal subspace constraint.

By substituting  constraint (\ref{CONST_V_s})
into the objective function  in (\ref{OBJ_P3}), we have
\begin{eqnarray}
& &\text{tr}(\mathbf{\Sigma} \textbf{V}^H \textbf{M}^{-1} \textbf{V}) \nonumber\\
&= &\text{tr}(\mathbf{\Sigma} \textbf{V}^H (\textbf{I}_m - \textbf{V}_2^H \textbf{V}_2)^H \textbf{M}^{-1} (\textbf{I}_m - \textbf{V}_2^H \textbf{V}_2) \textbf{V}) \nonumber\\
&= &\text{tr}(\mathbf{\Sigma} \textbf{V}^H \textbf{R} \textbf{V})
\label{TEMP_2}
\end{eqnarray}
where $\textbf{R} \triangleq (\textbf{I}_m - \textbf{V}_2^H \textbf{V}_2)^H \textbf{M}^{-1} (\textbf{I}_m - \textbf{V}_2^H \textbf{V}_2)$.

It can be easily shown that \(\text{rank}(\textbf{R})=\text{rank} (\textbf{I}_m - \textbf{V}_2^H \textbf{V}_2) = m-q\),
$\textbf{R}$ can be decomposed as
\begin{equation}
\textbf{R} = \textbf{V}_\textbf{R} \mathbf{\Lambda}_\textbf{R} \textbf{V}^H_\textbf{R}
\end{equation}
where $\textbf{V}_\textbf{R} \in St(m,m-q)$, $\mathbf{\Lambda}_\textbf{R}$ is an $(m-q) \times (m-q)$ diagonal
positive definite matrix with entries in non-decreasing order.
Since $\textbf{V}_2 \textbf{R} = \textbf{0}$, we have $\textbf{V}_2
\textbf{V}_\textbf{R} = \textbf{0}$. With $\textbf{V}_\textbf{R}$, the optimal Stiefel manifold matrix $\textbf{V}$ is given in the following theorem.

\begin{theorem}
\label{THEOREM_ZFBF_OPT_VS}
\emph{For the ZFB problem (P4), the solution is}
\end{theorem}
\begin{equation}
\textbf{V}^\ast = \textbf{V}_\textbf{R}
\left[
\begin{array}{c}
\textbf{I}_d\\
\textbf{0} \\
\end{array}
\right]_{(m-q) \times d} \nonumber.
\end{equation}

\subsubsection{Proof of Theorem \ref{THEOREM_ZFBF_OPT_VS}}
We first consider the case of $d = m-q$. In this case, $\textbf{V}$
is an $m \times (m-q)$ matrix on the Stiefel manifold.
Since $\textbf{V}$ and $\textbf{V}_\textbf{R}$ are of the same size
and perpendicular to \(\textbf{V}_2\), \(\textbf{V}\) and \(\textbf{V}_\textbf{R}\) are in the same signal subspace and there exists a unitary matrix \(\textbf{Q}\) that
satisfies
\begin{equation}
\label{TEMP_8}
\textbf{V} = \textbf{V}_\textbf{R} \textbf{Q} .
\end{equation}

By substituting
(\ref{TEMP_8}) into (\ref{FORM_T_S}), the transmit beamforming
matrix is given by
\begin{equation}
\label{TEMP_9}
\textbf{T} = \textbf{M}^{-1/2} \textbf{V}_\textbf{R} \textbf{Q} \mathbf{\Sigma}^{1/2} .
\end{equation}
Therefore finding the optimal $\textbf{T}$ is equivalent to
finding the optimal $\textbf{Q}$.
By substituting \eqref{TEMP_2} and (\ref{TEMP_8}) into (P4), the problem can be reformulated as
\begin{eqnarray}
\label{OBJ_P5}
\min_{\textbf{Q} \in St(m-q, m-q)}&& \text{tr}(\mathbf{\Sigma} \textbf{Q}^H \mathbf{\Lambda}_\textbf{R} \textbf{Q}).
\end{eqnarray}

By applying Lemma \ref{THEOREM_MIN_TRACE_PRODUCT} and considering the structure of $\mathbf{\Lambda}_\textbf{R}$, the optimal $\textbf{Q}$ can be easily constructed as $\textbf{Q}^\ast=
\textbf{I}_{m-q}$.

If $d < m-q$, we can treat this case as if there are still $m-q$ secondary data streams, among which $m-q - d$ data streams have zero-valued SNR requirements. In other words, the SNR constraint matrix is still an $(m-q) \times (m-q)$ diagonal matrix, but given by
$\mathrm{diag}(\rho_1,\cdots,\rho_d,0,\cdots,0)$.
The corresponding $\textbf{V}$ is still an $m \times (m-q)$ matrix on the Stiefel manifold. As a result, the result for the case of $d=m-q$ can be directly applied
and the optimal transmit beamforming matrix can be shown as
\begin{eqnarray}
\textbf{T}^\ast&=&\textbf{M}^{-1/2} \textbf{V}_\textbf{R}
\textbf{I}_{m-q} \mathrm{diag}(\rho_1^{1/2},...,\rho_d^{1/2},0,\cdots,0) \nonumber \\
&=&\textbf{M}^{-1/2} \textbf{V}_\textbf{R}
\left[
\begin{array}{c}
\textbf{I}_d \\
\textbf{0} \\
\end{array}
\right]_{(m-q) \times d}
\mathbf{\Sigma}^{1/2} .
\label{TEMP_7}
\end{eqnarray}

This completes the proof for Theorem \ref{THEOREM_ZFBF_OPT_VS}.

\emph{Remarks:} In (\ref{TEMP_7}), \(\mathbf{\Sigma}^{1/2}\) is used to allocate
power to the secondary data streams, \(\textbf{V}_\textbf{R}\) projects
the data streams to suitable signal subspaces in order to avoid interference at PR,
\(\textbf{Q}^* = [\textbf{I}_d {\ }\textbf{0}]^T\) is used to select the optimal dimension among the possible ones, and \(\textbf{M}^{-1/2}\) is used to handle the interference together with noise at SR.

\subsubsection{Other Choices of ${\textbf{Q}^\ast}$}
When the number of distinct SNR requirements is less than $d$, there
exist multiple data streams that have identical SNR requirement, and therefore
there will be multiple possible forms of $\textbf{Q}^{\ast}$ as well as
$\textbf{V}^\ast$. Suppose
\(\mathbf{\Sigma}\) has \(K\) distinct SNR values with
\begin{equation}
\mathbf{\Sigma} = \mathrm{diag}(\underbrace{\rho_1^{\prime},
\cdots, \rho _1^{\prime}}_{n_1}, \underbrace{\rho_2^{\prime},
\cdots, \rho_2^{\prime}}_{n_2}, \cdots, \underbrace{\rho_K^{\prime}, \cdots, \rho_K^{\prime}}_{n_K}) \nonumber
\end{equation}
where \(\sum_{k=1}^K n_k = d\) and
\(\rho_1^{\prime} > \rho_2^{\prime} \cdots > \rho_{n_K}^{\prime}\).
It turns out that the optimal $\textbf{Q}$ is a block
diagonal matrix in the form of
\begin{equation}
\textbf{Q}^* =
\left[
\begin{array}{c}
\mathrm{diag}(\textbf{Q}_1^{\ast},\cdots,\textbf{Q}_K^{\ast}) \\
\textbf{0}
\end{array}
\right] \in St(m-q,d)
\end{equation}
where
$\textbf{Q}_k^{\ast}$, $k=1,...,K$, can be any $n_k \times n_k$ unitary matrix.
For example, if $d = m-q=4$ and
$\mathbf{\Sigma} = \mathrm{diag}(a, a, b, b)$ ($a>b>0$),
the optimal \(\textbf{Q}^\ast\) is given by
\begin{equation}
\textbf{Q}^\ast =
\left[
\begin{array}{cc}
\textbf{Q}_{1}^{\ast} & \textbf{0}\\
\textbf{0} & \textbf{Q}_{2}^{\ast}
\end{array}
\right] \nonumber
\end{equation}
where $\textbf{Q}_1^{\ast}$ and $\textbf{Q}_2^{\ast}$
can be any $2 \times 2$ unitary matrices. 
Specifically, if all \(\rho_i\)'s are distinct,
\(\textbf{Q}^\ast\) must be a diagonal matrix with entries
of 1 or \(e^{j\theta}\); if all $\rho_i$'s are identical,
$\textbf{Q}^\ast$ can be an arbitrary unitary matrix.

\subsection{Nonzero Forcing Beamforming}
In the following, we focus on the NFB problem (P2) with $\xi >0$ and solve it by
examining its Lagrangian dual. Although problem (P2) is nonconvex, we provide a
sufficient condition for the strong duality to hold, and therefore general
convex optimization algorithms can be applied to solve the problem.

\subsubsection{Dual Problem}
As compared to ZFB, NFB relaxes the interference constraint at the PR by allowing a nonzero $\xi$, therefore it is possible for the secondary system to use fewer secondary transmit antennas or transmit more secondary data streams. Moreover, it provides more opportunities for the secondary system to access the channel.

The Lagrangian of (P2) is defined as
\begin{eqnarray}
\label{LAGRANGE_P3}
L(\textbf{V}, y) & \triangleq & \text{tr}(\mathbf{\Sigma} \textbf{V}^H
\textbf{M}^{-1} \textbf{V}) + y (\text{tr}(\mathbf{\Sigma} \textbf{V}^H
\textbf{M}_\text{x} \textbf{V}) - \xi) \nonumber \\
&=& \text{tr}(\mathbf{\Sigma} \textbf{V}^H (
\textbf{M}^{-1} + y \textbf{M}_\text{x})\textbf{V})  - y\xi)
\end{eqnarray}
where $y \geq 0$ is the Lagrange multiplier. Define the dual objective $g(\lambda)$ as an unconstrained
minimization of the Lagrangian
\begin{equation}
\label{DUAL_FUNCTION}
g(y) = \min_{\textbf{V} \in St(m,d)} L(\textbf{V}, y).
\end{equation}

Let $\lambda_i$, $i=1,...,m$, denote the $i$-th eigenvalue of $\textbf{M}^{-1} + y
\textbf{M}_\text{x}$ with ${\lambda}_1
\leq \cdots \leq {\lambda}_m$. Given a $y$, we can derive from Lemma \ref{THEOREM_MIN_TRACE_PRODUCT} that $L(\textbf{V}, y)$ is minimized
when $\textbf{V}$ is constructed as
\begin{equation}
\label{OPT_VS}
{\textbf{V}}_y = [\textbf{v}_1, \cdots,
\textbf{v}_d]
\end{equation}
where $\textbf{v}_i$, $i=1,...,d$, denotes the eigenvector of $\textbf{M}^{-1} + y
\textbf{M}_\text{x}$ corresponding to ${\lambda}_i$. Therefore, matrix ${\textbf{V}}_y$, as well as the corresponding
sum transmit power $\text{tr}(\mathbf{\Sigma} {\textbf{V}}_y^H \textbf{M}^{-1} {\textbf{V}}_y)$
and interference power $\text{tr}(\mathbf{\Sigma} {\textbf{V}}_y^H \textbf{M}_\text{x}
{\textbf{V}}_y)$, are functions of $y$. The Lagrange dual problem associated with (P2) is given by
\begin{subequations}
\begin{eqnarray}
\!\!\!\!\!\!\!\!\!\!\!\!\!\!\!\!\!\!\!\!\!\!\!\!\!\!\!\!\!\!\!\!\!\!\!\!\!\!\!\!\!\!\!\!\!\!\!\!\!\!\!\!(\text{P5}):\ \ \ \ \ \ \ \ \ \ \ \ \ \ \ \ \ \ \max && g(y) \\
\mathrm{s. t.} && y \geq 0.
\end{eqnarray}
\end{subequations}

Note that the original problem (P2) is nonconvex because the domain of $\textbf{V}$ is nonconvex.
Therefore, we cannot directly tell if the strong duality holds or not. The following theorem
provides a sufficient condition for strong duality between the primal problem (P2) and the dual problem (P5).

\begin{theorem}
\label{THEOREM_STRONG_DUALITY}
The strong duality between (P2) and (P5) holds if the feasibility test is passed,
i.e., $\xi \geq \xi_0$.
\end{theorem}

\emph{Remarks:} The feasible domain is nonempty if and only if $\xi \geq \xi_0$. As a
result, as long as the ST can access the channel, we can always solve the dual problem
to obtain the optimal beamforming solution, which is of zero duality gap. 

\subsubsection{Proof of Theorem \ref{THEOREM_STRONG_DUALITY}}
To the best of our knowledge, although there are a number of constraint qualifications under
which strong duality holds even if the primal problem is nonconvex \cite{REF_BOYD_CVXOPT, BNO, OB}, none of them can be directly
applied to decide if the strong duality holds in our case.

Define $\epsilon(y) \triangleq y (\text{tr}(\mathbf{\Sigma} {\textbf{V}}_y^H
\textbf{M}_\text{x} {\textbf{V}}_y) - \xi)$. Note that for an arbitrary $y$, $\epsilon(y)$ represents the difference between $g(y)$ and the corresponding primal objective function value $\text{tr}(\mathbf{\Sigma} {\textbf{V}}_y^H \textbf{M}^{-1} {\textbf{V}}_y)$.
It is known that the dual function $g(y)$ yields a lower bound of (P2) for all $y \geq 0$. If there exists a $y$ such that $\epsilon(y) = 0$, then the result of the dual problem (P5) becomes $\text{tr}(\mathbf{\Sigma} {\textbf{V}}_y^H \textbf{M}^{-1} {\textbf{V}}_y)$, which is exactly equal to the corresponding primal objective function value, and serves as the infimum of (P2). That is, $y$ and $\textbf{V}_y$ are the optimal solution of the dual problem (P5) and the primal problem (P2), respectively, and therefore zero duality gap holds. Otherwise, if no $y$ and the corresponding $\textbf{V}_y$ make $\epsilon(y)=0$, we cannot find the optimal  $\textbf{V}$ by solving the dual problem (P5). As a result, $g(y)$ becomes a loose lower bound
of (P2) for all $y\geq 0$, which implies that the zero duality gap does not hold.

We will need the following lemma to prove Theorem \ref{THEOREM_STRONG_DUALITY}.
\begin{lemma}
\label{LEMMA_3}
The transmit power $\text{tr}(\mathbf{\Sigma} {\textbf{V}}_y^{H}
\textbf{M}^{-1} {\textbf{V}}_y)$ is a monotonically increasing
function of $y$ and the interference power $\text{tr}(\mathbf{\Sigma}
{\textbf{V}}_y^{H} \textbf{M}_\text{x} {\textbf{V}}_y)$
is a monotonically decreasing function of $y$.
\end{lemma}

\begin{IEEEproof}
Please refer to Appendix D.
\end{IEEEproof}

Lemma \ref{LEMMA_3} suggests that the minimum interference power is achieved at $y = \infty$. If the secondary transmission is feasible, i.e., $\xi \geq \xi_0$, it is easy to show that
either of the following two conditions is satisfied:
\begin{eqnarray}
\left\{
\begin{array}{cc}
\text{tr}(\mathbf{\Sigma}
{\textbf{V}}_y^{H} \textbf{M}_\text{x} {\textbf{V}}_y) = \xi, & \exists y > 0 \nonumber \\
\text{tr}(\mathbf{\Sigma}
{\textbf{V}}_y^{H} \textbf{M}_\text{x} {\textbf{V}}_y) \leq \xi, & y = 0. \nonumber
\end{array}
\right.
\end{eqnarray}
Therefore, $\epsilon(y)=0$
is always satisfied, i.e., the strong duality holds. That is to say, passing the
feasibility test is the sufficient condition for zero duality gap in our problem.
Furthermore, the ${\textbf{V}}_y$
corresponding to the minimum feasible $y$ of the dual problem
is the optimal solution that minimizes the sum transmit power. General convex optimization algorithms can be applied to find the optimal $y$ and consequently the optimal transmit beamforming matrix.

\begin{figure}[!t]
\centering
  \includegraphics[width=4.9in]{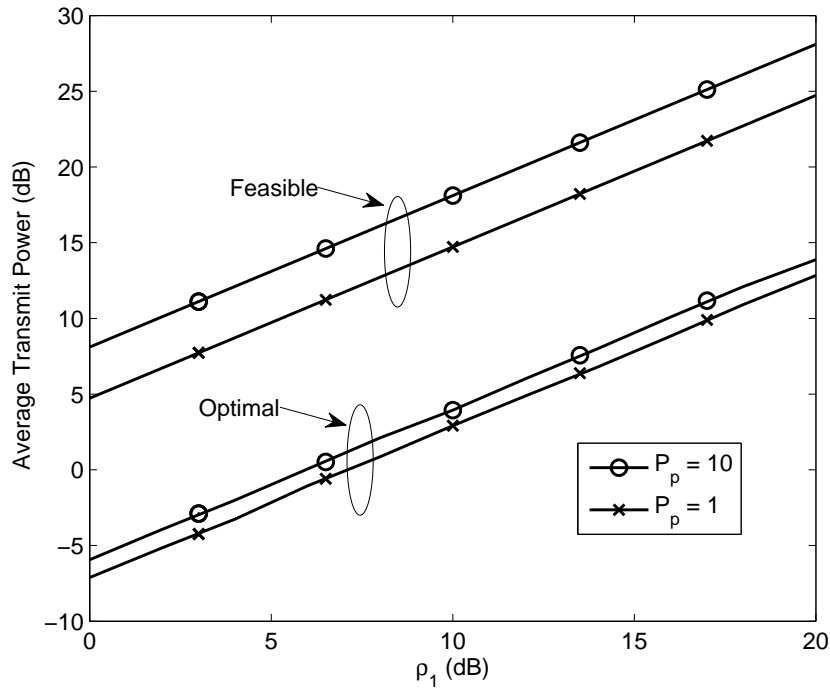}
  \caption{Optimal beamforming versus feasible beamforming. $d=1$.}
  \label{fig_2}
\end{figure}

\begin{figure}[!t]
\centering
  \includegraphics[width=4.9in]{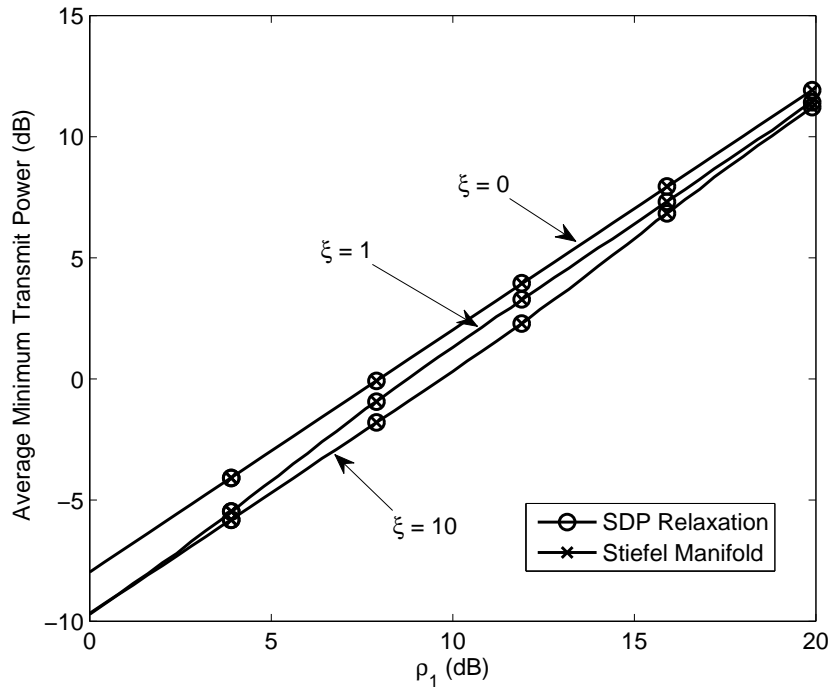}
  \caption{Impact of the secondary SNR requirement. $d=1$.}
  \label{fig_3}
\end{figure}

\section{Numerical Results}

Numerical results are provided in this section to illustrate the performance of the proposed beamforming solutions under various channel conditions and system requirements.
We assume that each entry of channel matrices $\textbf{G}_\text{x}$,
$\textbf{H}_\text{x}$, and $\textbf{H}$ is
an i.i.d. random variable, distributed as $\mathcal{CN}(0,1)$. The primary transmit power ${P}_{p}$ is assumed to be 1, unless otherwise specified. Throughout this section, we set $p=q=2$ and $m=n=5$, and $d$ is selected to satisfy $m-q \geq d$ such that secondary access is always possible. Monte Carlo simulations with 2000 randomly generated channel-groups ($\textbf{H}, \textbf{H}_\text{x}, \textbf{G}_\text{x}$) are implemented, and the \emph{average} minimum secondary transmit power is plotted versus the secondary per data stream SNR constraint or the primary interference power constraint in Figs. 2-8. In the case of multiple secondary data streams (i.e., $d>1$), equal SNR requirements are considered in all the figures except Fig.~8.

Firstly, we consider the case of single secondary data stream in Figs. 2 and 3. Fig.~\ref{fig_2} shows the minimum required average secondary transmit power based on the derived optimal beamforming matrix for ZFB. As a comparison, the required secondary transmit power
averaged over all feasible beamforming matrices is also shown in Fig.~\ref{fig_2}. Here,
the feasible beamforming matrix refers to any beamforming matrix that satisfies
the interference constraint in (3b) and the SNR requirement in
(3c). From (5), it is clear that the required secondary transmit power linearly increases with the secondary SNR requirement, as shown in Fig.~\ref{fig_2}. In Fig.~\ref{fig_2}, we observe that the optimal beamforming matrix brings
significant power saving over the feasible beamforming matrix and that this power saving increases with the primary transmit power.

In Fig.~\ref{fig_3} we compare the SDP relaxation approach in Section III and the Stiefel manifold transformation approach in Section IV for both ZFB and NFB cases. Fig.~\ref{fig_3} verifies that the SDP relaxation and the Stiefel manifold transformation approach achieve the same optimal performance, as expected.

\begin{figure}[!t]
\centering
  \includegraphics[width=4.9in]{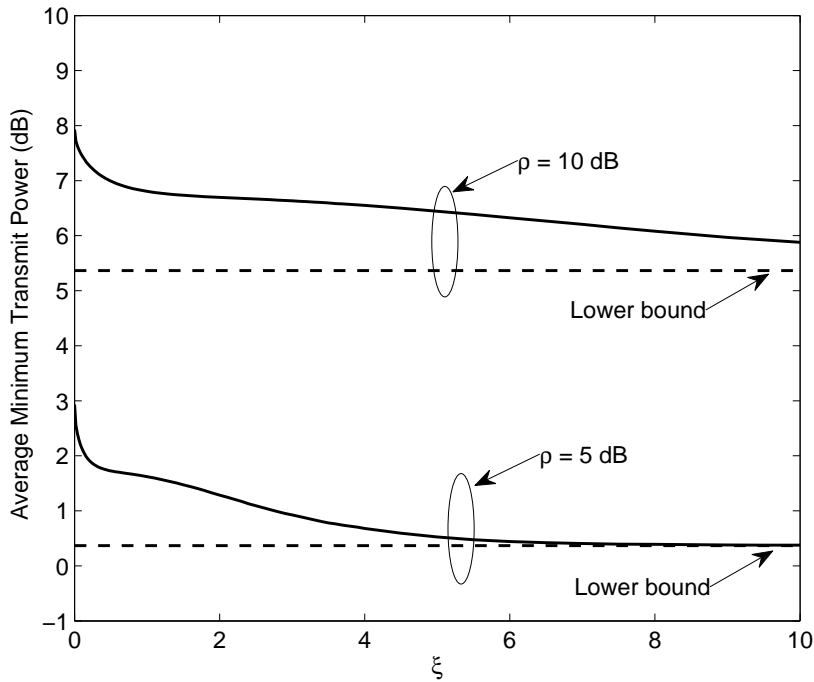}
  \caption{Impact of the primary interference constraint. $d=2$ and $\rho_1=\rho_2=\rho$.}
  \label{fig_4}
\end{figure}

\begin{figure}[!t]
\centering
  \includegraphics[width=4.9in]{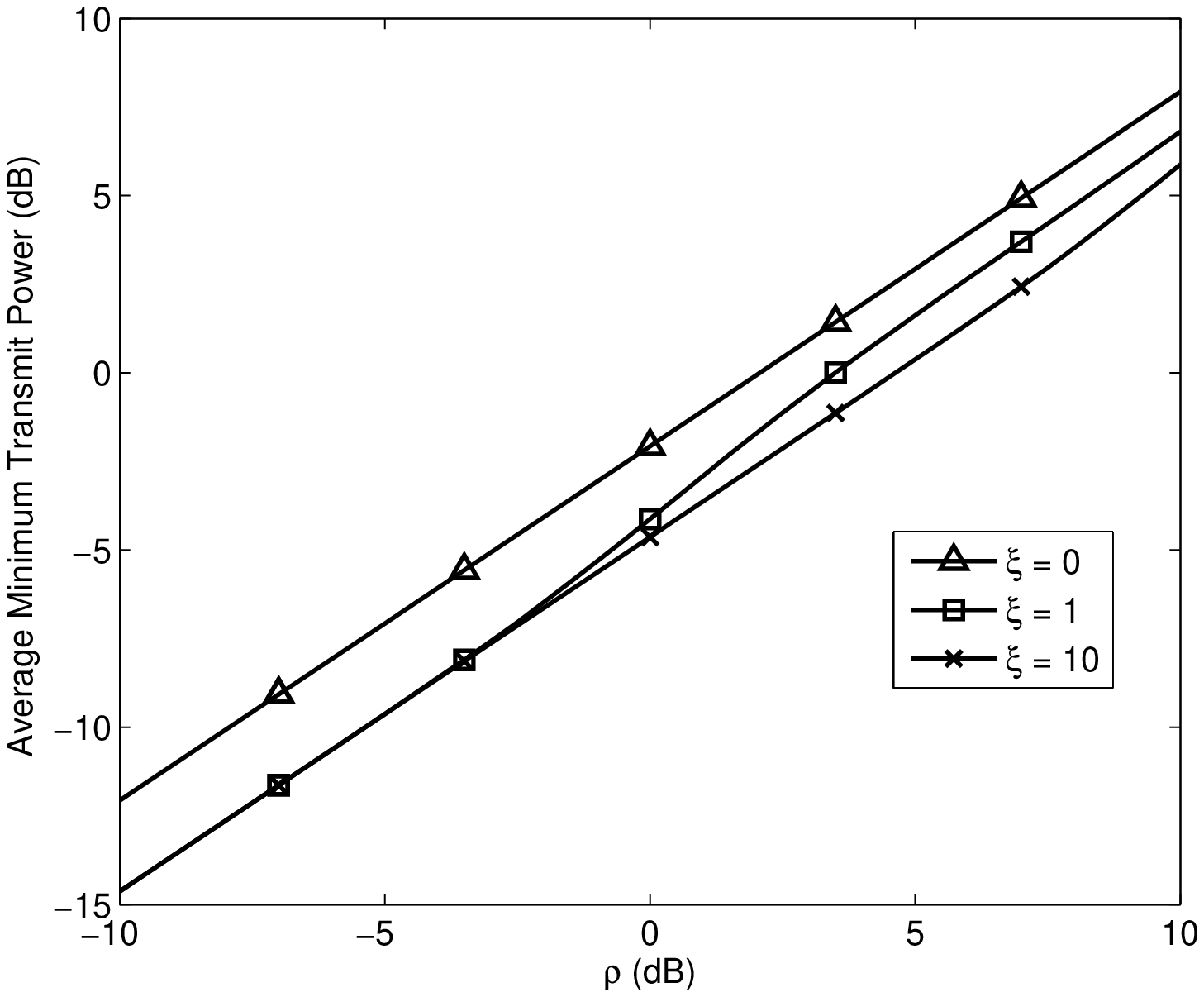}
  \caption{Impact of the secondary SNR requirement. $d=2$ and $\rho_1=\rho_2=\rho$.}
  \label{fig_5}
\end{figure}

Next, we consider the case of multiple secondary data streams. For simplicity, only the optimal beamforming solutions are presented in the rest of this section. The impact of interference power constraint $\xi$ and secondary per data stream SNR requirement $\rho$ on the minimum required secondary transmit power is shown in Figs.~\ref{fig_4} and \ref{fig_5}, where $d=2$ secondary data streams are considered. From Fig.~\ref{fig_4}, it is observed that for a fixed $\rho$,
the required transmit power decreases significantly when $\xi$ is slightly increased over the zero value (the beamforming is changed from ZFB to NFB) while the required power decreases slowly when the positive $\xi$ further increases. This is not unexpected since ZFB significantly restricts the available beamforming dimensions as compared to NFB. In this case, there are only three dimensions available for ZFB and the other two dimensions are not permitted to use due to the nature of ZFB, while NFB can use all the five dimensions for beamforming. On the other hand, for both ZFB and NFB (i.e., zero and nonzero $\xi$ values), a lower secondary SNR requirement always brings a significant reduction in the required secondary transmit power. The lower bound of the secondary transmit power when $\xi \rightarrow \infty$ (it can be easily obtained by dropping constraint (3b) in (P1)) is also plotted in Fig.~\ref{fig_4}, where we observe that $\rho$ is more dominant than $\xi$ in determining the required secondary power. The similar finding can be observed in Fig.~\ref{fig_5}, as well as the relatively small difference among the nonzero $\xi$ values. In particular, we observe  that at small $\rho$ values, there is almost no difference between the nonzero $\xi$ values in Fig.~\ref{fig_5}. This is because when the secondary SNR requirement is small, the required secondary power is small and thus the interference caused to the primary receiver is also small and always below the underlying primary interference constraint $\xi$. When $\rho$ increases, the required secondary transmit power needs to increase and therefore the caused interference may no longer be lower than $\xi$. For a more strict interference constraint (i.e., a lower $\xi$), the available secondary beamforming directions are subject to more restrictions and therefore more secondary transmit power is needed to satisfy the primary interference constraint as well as the secondary per data stream SNR constraint at the same time.

\begin{figure}[!t]
\centering
  \includegraphics[width=4.9in]{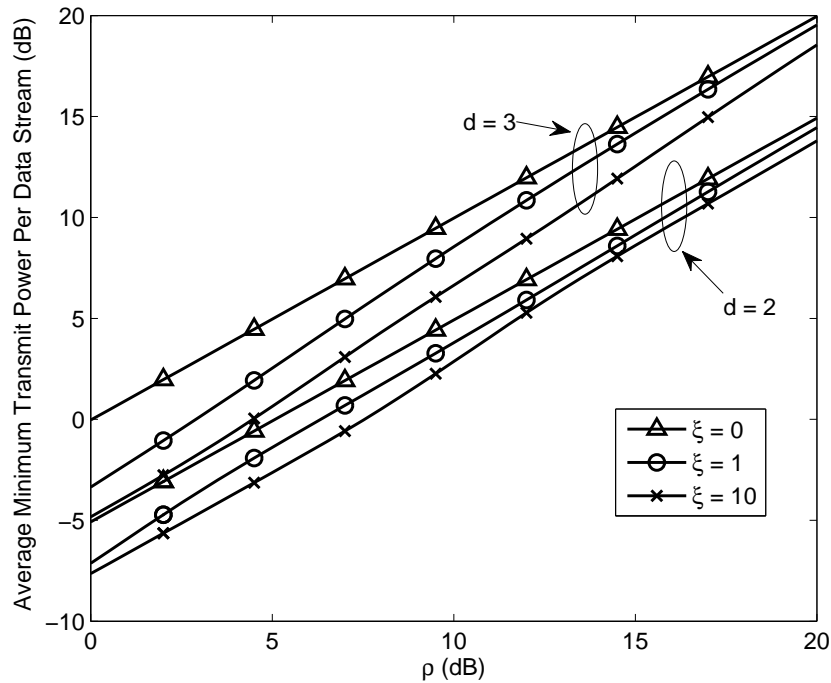}
  \caption{Minimum required transmit power per secondary data stream.}
  \label{fig_6}
\end{figure}

\begin{figure}[!t]
\centering
  \includegraphics[width=4.9in]{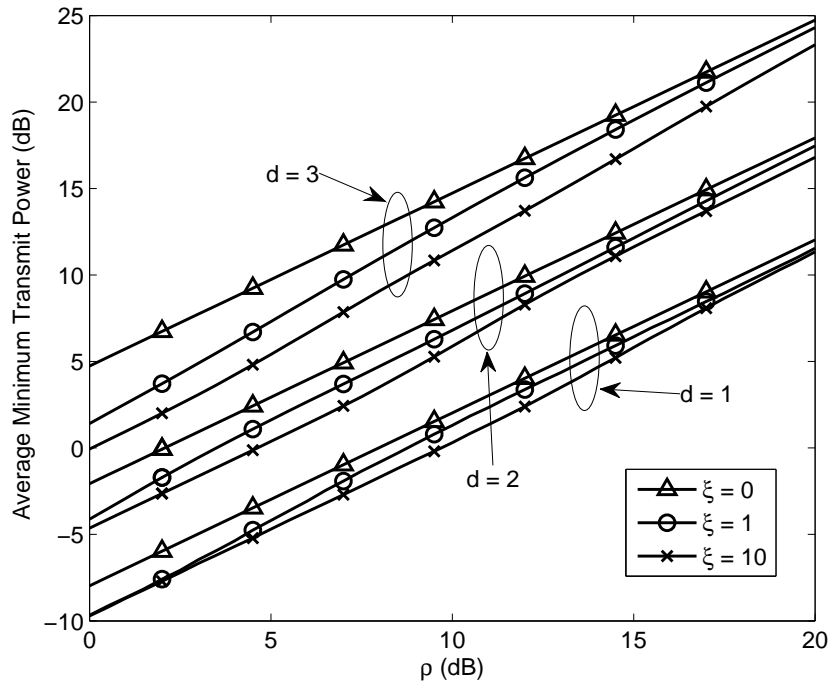}
  \caption{Impact of the number of secondary data streams.}
  \label{fig_7}
\end{figure}

The minimum required secondary power per data stream is investigated in Fig.~\ref{fig_6}, where the SNR requirement of each secondary data stream is assumed to be equal to each other, denoted as $\rho$. Together with Fig.~\ref{fig_3}, it is seen that when $d$ is increased from 1 to 3, more power needs to be allocated to each data stream on average. When $d$ is small, ST can choose
channels with good conditions to transmit data streams and therefore the required power per data stream can be relatively small. As $d$ increases, while satisfying
the primary interference power constraint as a higher priority, ST may have to use channels with poor conditions and hence more power is required to transmit data streams with the same secondary SNR requirement per data stream. Based on Figs.~\ref{fig_3} and \ref{fig_6}, the total minimum required transmit power with different number of secondary data streams is shown in Fig.~\ref{fig_7}.

\begin{figure}[!t]
\centering
  \includegraphics[width=4.9in]{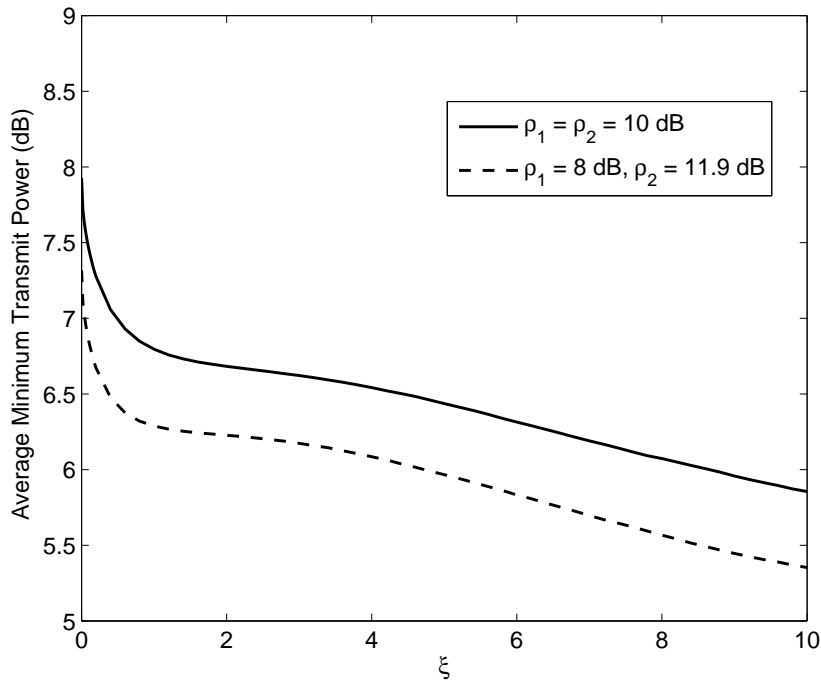}
  \caption{Distinct versus identical secondary SNR requirements.}
  \label{fig_8}
\end{figure}

Fig.~\ref{fig_8} illustrates the impact of distinct and identical
secondary SNR requirements for both ZFB and NFB (zero and nonzero $\xi$ values) cases with $d=2$. Here, the SNR requirements have been chosen such
that the achievable sum-rate is the same for both distinct and identical SNR requirements. Fig.~\ref{fig_8} shows that the case of distinct SNR requirements costs
less power than the identical SNR requirements case. The reason is given as follows. MIMO provides parallel channels for the transmission of multiple data streams but the underlying conditions of different channels may not be the same. From the secondary system point of view, with distinct SNR requirements,
the secondary data stream with a higher SNR requirement can be allocated to a channel
with better conditions to save the transmit power. This is why the case of distinct
SNR requirements outperforms the case of identical SNR requirements in terms of lower secondary transmit power.

\section{Conclusion}
Cognitive beamforming for multiple secondary data streams is considered in this paper. The optimal beamforming strategy is designed
to minimize the secondary transmit power consumption, subject to the individual secondary per data stream SNR constraints
and primary interference power constraint. By exploiting the Stiefel manifold, we derived
the closed form solution for zero forcing beamforming; while for nonzero forcing beamforming, we prove that the strong duality holds for the nonconvex primal problem and thus the optimal solution can be easily obtained by solving the dual problem. As for the future work, we are interested in exploring the impact of imperfect channel state information at secondary users and designing robust cognitive beamforming solutions.

\appendices

\section{}

Letting $\textbf{T} = [{\textbf{t}}_1,\cdots,{\textbf{t}}_d]$ and denoting ${\textbf{r}}_i$ ($i=1,...,d$) as the receive beamforming vector for the $i$th data stream, the SNR for the $i$th data stream is given by
\begin{equation}
\label{RCV_BF}
\hat{\rho}_i
= \frac{{\textbf{r}}_i \textbf{H} {\textbf{t}}_i {\textbf{t}}_i^H \textbf{H}^H {\textbf{r}}_i^H}
{{\textbf{r}}_i (\sum_{k \neq i} \textbf{H} {\textbf{t}}_k {\textbf{t}}_k^H \textbf{H}^H  +  \textbf{W}) {\textbf{r}}_i^H}
=\frac{{\textbf{r}}_i \textbf{M}_1 {\textbf{r}}_i^H}{{\textbf{r}}_i \textbf{M}_2 {\textbf{r}}_i^H}.
\end{equation}
Eqn. \eqref{RCV_BF} is a generalized Rayleigh quotient and the ${\textbf{r}}_i$ that maximizes $\hat{\rho}_i$ is the dominant eigenvector of
$(\textbf{M}_2^{-1/2})^H \textbf{M}_1 \textbf{M}_2^{-1/2}$. It thus follows that the maximum $\hat{\rho}_i$ is equal to the $i$-th eigenvalue of $\textbf{T}^H
\textbf{H}^H \textbf{W}^{-1} \textbf{H} \textbf{T}$.

\section{}

For any $\textbf{T}$ that satisfies
(\ref{eq3}), we define $\textbf{P} \triangleq
\textbf{M}^{1/2} \textbf{T} \in \mathbb{C}^{m \times d}$ and let $\textbf{p}_i$ denote
the $i$th column of $\textbf{P}$. Since $\mathbf{\Sigma}$ is a diagonal matrix, it follows from \eqref{eq3} that all $\textbf{p}_i$'s are orthogonal to each other and the norm of $\textbf{p}_i$ is
equal to ${\Sigma}_{ii}^{1/2}$. We normalize each $\textbf{p}_i$
to form a matrix $\textbf{V} = [\textbf{p}_1/|\textbf{p}_1| \cdots,
\textbf{p}_d/|\textbf{p}_d|] = \textbf{P} \mathbf{\Sigma}^{-1/2} \in St(m,d)$. This
directly yields $\textbf{T} = \textbf{M}^{-1/2} \textbf{P} = \textbf{M}^{-1/2} \textbf{V}
\mathbf{\Sigma}^{1/2}$.

\section{}

It is shown in \cite{REF_LASSERRE_TRACE_INEQUALITY} that for any $w \times w$ Hermitian matrices
$\textbf{A}$ and $\textbf{B}$,
\begin{equation}
\label{TRACE_INEQUALITY}
\sum_{i = 1}^w \lambda_{i}(\textbf{A}) \lambda_{w-i+1} (\textbf{B})
\leq \text{tr}(\textbf{A} \textbf{B})
\end{equation}
where $\lambda_i(\textbf{X})$ denotes the $i$-th eigenvalue of matrix $\textbf{X} \in
\mathbb{C}^{w \times w}$ and  $\lambda_1(\textbf{X}) \geq \cdots \geq \lambda_w(\textbf{X})$.

Let $\mathbf{\Delta}^\prime \triangleq \mathrm{diag} (\mathbf{\Delta} {\ } \textbf{0})
\in \mathbb{C}^{v \times v}$ and $\mathbf{\Theta}^\prime \triangleq
[\mathbf{\Theta} {\ } \mathbf{\Theta}_\perp] \in \mathbb{C}^{v \times v}$
where $\text{rank}(\mathbf{\Theta}_\perp) = v-u$ and $\mathbf{\Theta}_\perp^H
\mathbf{\Theta} = \textbf{0}$. Note that $\lambda_{u+1}(\mathbf{\Delta}^\prime) = \cdots =
\lambda_{v}(\mathbf{\Delta}^\prime) = 0$ and the eigenvalues of $\mathbf{\Theta}^{\prime H}
\mathbf{\Omega} \mathbf{\Theta}^\prime$ are equal to those of $\mathbf{\Omega}$. By
using \eqref{TRACE_INEQUALITY}, we get
\begin{eqnarray}
\sum_{i = 1}^u \delta_i \omega_{v-i+1} &=&
\sum_{i = 1}^v \lambda_{i}(\mathbf{\Delta}^\prime) \lambda_{v-i+1} (\mathbf{\Omega}) \nonumber \\
&\leq & \text{tr}(\mathbf{\Delta}^\prime \mathbf{\Theta}^{\prime H}
\mathbf{\Omega} \mathbf{\Theta}^\prime) \nonumber \\
&=& \text{tr}(\mathbf{\Delta} \mathbf{\Theta}^H
\mathbf{\Omega} \mathbf{\Theta}). \nonumber
\end{eqnarray}
The equality holds when $\mathbf{\Theta}^{\prime H}
\mathbf{\Omega} \mathbf{\Theta}^\prime = \mathrm{diag}(\omega_v, \cdots,
\omega_1)$, i.e., $\mathbf{\Theta}^\prime = [\boldsymbol{\varphi}_v, \cdots, \boldsymbol{\varphi}_1]$. It thus follows that $\mathbf{\Theta} = [\boldsymbol{\varphi}_v, \cdots, \boldsymbol{\varphi}_{v-u+1}]$.
This completes the proof.

\section{}

Let $\mathbf{\Sigma}^\prime = \mathrm{diag}(\mathbf{\Sigma}, \textbf{0}) \in \mathbb{R}^{m \times m}$,
and $\textbf{V}^\prime = [{\textbf{V}}_y {\textbf{V}}_{y \perp}]$
where ${\textbf{V}}_{y \perp} \in \mathbb{C}^{m \times (m-d)}$ and its $i$-th $(i=1,...,m-d)$ column is the
eigenvector of $\textbf{M}^{-1} + y \textbf{M}_\text{x}$ corresponding to ${\lambda}_{d+i}$.
The dual problem (P5) can be equivalently rewritten as
\begin{equation}
\max_{y \geq 0} \min_{\textbf{V}^\prime \in St(m,m)}
\text{tr}(\mathbf{\Sigma}^\prime \textbf{V}^{\prime H}
(\textbf{M}^{-1}+ y \textbf{M}_\text{x})\textbf{V}^\prime - y\xi).
\end{equation}

Let $\pi: \mathbb{C}^{m \times m} \rightarrow St(m,m)$ be a projection operator mapping an $m \times m$
complex matrix to the closest point on the Stiefel manifold \cite{REF_MANTON}.
Suppose that $\textbf{Z}$ is the derivative of the $\textbf{V}^\prime$ with respect to
$y$. It means that given a sufficiently small increment $\Delta y$, the $\textbf{V}^\prime$
related to $\textbf{M}^{-1} + (y+\Delta y)\textbf{M}_\text{x}$, denoted by
$\textbf{V}^{\prime \prime}$, can be approximated as
\begin{equation}
\textbf{V}^{\prime \prime} \approx \pi(\textbf{V}^\prime + \Delta y \textbf{Z})
= \textbf{V}^\prime + \Delta y \textbf{Z} + O(\Delta y^2) \nonumber
\end{equation}
where $\textbf{Z}$ can be written as $\textbf{Z} = \textbf{V}^\prime
\textbf{C}$ and $\textbf{C} \in \mathbb{C}^{m \times m}$ is a skew-Hermitian
matrix ($\textbf{C}^H + \textbf{C} = \textbf{0}$) to be determined \cite{REF_MANTON}.
Let us define ${\mathbf{\Lambda}} \triangleq \mathrm{diag}({\lambda_1}, \cdots,
{\lambda}_m)$ and $\textbf{F} \triangleq \textbf{V}^{\prime H} \textbf{M}_\text{x}
\textbf{V}^\prime$.
Since the columns of $\textbf{V}^{\prime \prime}$ are the eigenvectors of
$\textbf{M}^{-1} + (y+\Delta y)\textbf{M}_\text{x}$, the expression $\textbf{V}^{\prime \prime H}
(\textbf{M}^{-1} + (y+\Delta y)\textbf{M}_\text{x}) \textbf{V}^{\prime \prime }$ is a diagonal matrix consisting of the eigenvalues of $\textbf{M}^{-1} + (y+\Delta y)\textbf{M}_\text{x}$.
$\textbf{V}^{\prime \prime H} (\textbf{M}^{-1} + (y+\Delta y)\textbf{M}_\text{x})
\textbf{V}^{\prime \prime }$ can be decomposed as
\begin{eqnarray}
\label{DERIV_OF_EIGENVALUE}
&&\textbf{V}^{\prime \prime H} (\textbf{M}^{-1} +
(y+\Delta y)\textbf{M}_\text{x}) \textbf{V}^{\prime \prime } \nonumber \\
&\approx&\pi(\textbf{V}^\prime + \Delta y \textbf{Z})^H (\textbf{M}^{-1} +
(y+\Delta y)\textbf{M}_\text{x}) \pi(\textbf{V}^\prime + \Delta y \textbf{Z}) \nonumber \\
&=& \textbf{V}^{\prime H} (\textbf{M}^{-1} + y \textbf{M}_\text{x}) \textbf{V}^\prime
+ \Delta y \left(\textbf{Z}^H (\textbf{M}^{-1} + y \textbf{M}_\text{x}) \textbf{V}^\prime \right. \nonumber \\  && \left. +
\textbf{V}^{\prime H} (\textbf{M}^{-1} + y \textbf{M}_\text{x}) \textbf{Z} + \textbf{V}^{\prime H}
\textbf{M}_\text{x} \textbf{V}^\prime \right) + O(\Delta y^2) \nonumber \\
&=& {\mathbf{\Lambda}} + \Delta y ( - \textbf{C} {\mathbf{\Lambda}} + {\mathbf{\Lambda}} \textbf{C} +
\textbf{F} )
+ O(\Delta y^2)
\end{eqnarray}
which indicates that matrix $( - \textbf{C} {\mathbf{\Lambda}} + {\mathbf{\Lambda}} \textbf{C} + \textbf{F})$ has to be diagonal.
By computing the expressions of the entries of $- \textbf{C} {\mathbf{\Lambda}} + {\mathbf{\Lambda}}
\textbf{C} + \textbf{F}$ and setting the off-diagonal entries to be zero, it turns out that
\begin{equation}
\label{OFFDIAG_C}
(\lambda_i - \lambda_j) C_{ij} = - F_{ij}, \forall i \neq j.
\end{equation}

Firstly, we consider the impact of positive $\Delta y$ on the interference power. Let
$\textbf{C}^{(d)}$ and $\textbf{C}^{(od)}$ denote the matrices that only contain the diagonal
and off-diagonal entries of $\textbf{C}$, respectively. With \eqref{OFFDIAG_C}, the interference power difference
at a sufficiently small increment $\Delta y > 0$ can be shown to be 
\begin{eqnarray}
\label{VIBRATION_Y_ON_INT_POWER}
&&\text{tr} (\mathbf{\Sigma}^\prime \pi(\textbf{V}^\prime\! +\! \Delta y \textbf{Z})^H
\textbf{M}_\text{x} \pi(\textbf{V}^{\prime H}\! +\! \Delta y \textbf{Z}))\! -\!
\text{tr} (\mathbf{\Sigma}^\prime \textbf{V}^{\prime H} \textbf{M}_\text{x}
\textbf{V}^\prime) \nonumber \\
&\approx& 2\Delta y \text{Re} \{\text{tr} (\mathbf{\Sigma}^\prime \textbf{V}^{\prime H} \textbf{M}_\text{x}
\textbf{Z})\} \nonumber \\
&=& 2\Delta y \text{Re} \{\text{tr} (\mathbf{\Sigma}^\prime \textbf{F} (\textbf{C}^{(d)} + \textbf{C}^{(od)}))\} \nonumber \\
&\aeq& 2\Delta y \text{Re} \{\text{tr} (\mathbf{\Sigma}^\prime \textbf{F} \textbf{C}^{(od)})\} \nonumber \\
&=& 2\Delta y \sum_{i = 1}^d \rho_i (\sum_{j \neq i} F_{ij} C^{(od)}_{ji}) \nonumber \\
&\beq& 2\Delta y \sum_{i = 1}^d \rho_i (\sum_{j \neq i} (\lambda_i - \lambda_j)  |C^{(od)}_{ij}|^2 ) \nonumber \\
&=& 2\Delta y \Big( \sum_{i = 1}^{d-1} \sum_{j = i+1}^{d} (\rho_i - \rho_j) (\lambda_i - \lambda_j) |C^{(od)}_{ij}|^2 + \sum_{i = 1}^d \sum_{j = d+1}^m \rho_i (\lambda_i - \lambda_j) |C^{(od)}_{ij}|^2 \Big) \nonumber \\
&\cleq & 0.
\end{eqnarray}
In \eqref{VIBRATION_Y_ON_INT_POWER},
(a) holds because $\text{Re} \{\text{tr} (\mathbf{\Sigma}^\prime \textbf{V}^{\prime H}
\textbf{M}_\text{x} \textbf{V}^\prime \textbf{C}^{(d)})\} = \text{Re} \{\text{tr}
((\textbf{C}^{(d)} \mathbf{\Sigma}^\prime) (\textbf{V}^{\prime H}
\textbf{M}_\text{x} \textbf{V}^\prime))\}=0$, where $\textbf{C}^{(d)} \mathbf{\Sigma}^\prime$
is skew-Hermitian and $\textbf{V}^{\prime H}
\textbf{M}_\text{x} \textbf{V}^\prime$ is Hermitian; 
(b) is obtained by substituting \eqref{OFFDIAG_C}; (c) is because $\rho_i \geq \rho_j$ and $\lambda_i \leq \lambda_j$
for $i < j$, and the equality holds if and only if $\rho_i = \rho_j$, $\lambda_i = \lambda_j$, $\forall
i \neq j$. Considering the random nature of wireless channels, it is nearly impossible for the equality under (c) to hold.
Therefore, we conclude that the interference power in the form of $\text{tr} (\mathbf{\Sigma}^\prime \textbf{V}^{\prime H} \textbf{M}_\text{x}
\textbf{V}^\prime)$
is a monotonically decreasing function of $y$.

Next, we consider how the value of $y$ affects the required secondary transmit power.
Eqn. \eqref{LAGRANGE_P3} can be rewritten as
\begin{equation}
\label{LAGRANGE_ALPHA}
L(\textbf{V}^\prime, y) = (1+y)(\text{tr}(\mathbf{\Sigma}^\prime \textbf{V}^{\prime H}
(\alpha \textbf{M}^{-1} + (1-\alpha) \textbf{M}_\text{x}) \textbf{V}^\prime)) - y \xi
\end{equation}
where $\alpha = 1/(1+y) \in [0,\infty)$. Note that the selected $\textbf{V}^\prime$ in the
dual problem is determined by the weighted sum of $\textbf{M}^{-1}$ and $\textbf{M}_\text{x}$.
A larger $y$ increases the weight of $\textbf{M}_\text{x}$ and at the same time
decreases the weight of $\textbf{M}^{-1}$. Let $\Delta \alpha$ be the change of $\alpha$ as
$y$ becomes $y+\Delta y$ and $\textbf{E}\triangleq \textbf{V}^{\prime H}
\textbf{M}^{-1} \textbf{V}^\prime$. Similar to \eqref{DERIV_OF_EIGENVALUE}, it can be derived that
\begin{eqnarray}
&&\textbf{V}^{\prime \prime H} ((\alpha + \Delta \alpha)\textbf{M}^{-1} +
(1-\alpha-\Delta \alpha)\textbf{M}_\text{x}) \textbf{V}^{\prime \prime } \nonumber \\
&=& \alpha {\mathbf{\Lambda}} - \frac{\Delta \alpha}{\alpha} [( - \textbf{C} {\mathbf{\Lambda}}
+ {\mathbf{\Lambda}} \textbf{C}) -
\alpha (\textbf{E} - \textbf{F})] + O(\Delta \alpha^2) \nonumber \\
&=& \alpha {\mathbf{\Lambda}} - {\Delta \alpha} [ \alpha (- \textbf{C} {\mathbf{\Lambda}} +
{\mathbf{\Lambda}} \textbf{C} + \textbf{F}) + (1-\alpha) (- \textbf{C} {\mathbf{\Lambda}} +
{\mathbf{\Lambda}} \textbf{C} - {\alpha}\textbf{E}/(1-\alpha))]/\alpha
+ O(\Delta \alpha^2)
\end{eqnarray}
where matrix $(- \textbf{C} {\mathbf{\Lambda}} +
{\mathbf{\Lambda}} \textbf{C} - \alpha \textbf{E}/(1-\alpha))$
should be diagonal. It thus follows that
\begin{equation}
\label{eq42}
(\lambda_i - \lambda_j) C_{ij} = \alpha E_{ij} /(1-\alpha)=  E_{ij}/y.
\end{equation}

With \eqref{eq42}, the difference of the required transmit power at $\Delta y > 0$ can be shown to be
\begin{eqnarray}
\label{VIBRATION_Y_ON_TX_POWER}
&& \text{tr} (\mathbf{\Sigma}^\prime \pi(\textbf{V}^\prime + t \textbf{Z})^H
\textbf{M}^{-1} \pi(\textbf{V}^{\prime H} + t \textbf{Z})) -
\text{tr} (\mathbf{\Sigma}^\prime \textbf{V}^{\prime H} \textbf{M}^{-1}
\textbf{V}^\prime) \nonumber \\
&\approx & - \frac{2\Delta y}{y} \Big( \sum_{i = 1}^{d-1} \sum_{j = i+1}^{d} (\rho_i - \rho_j)
(\lambda_i - \lambda_j) |C^{(od)}_{ij}|^2  + \sum_{i = 1}^d \sum_{j = d+1}^m \rho_i (\lambda_i -
\lambda_j) |C^{(od)}_{ij}|^2 \Big) \nonumber \\
&\geq & 0.
\end{eqnarray}

With a similar argument used for \eqref{VIBRATION_Y_ON_INT_POWER}, we conclude that the
required secondary transmit power in terms of $\text{tr} (\mathbf{\Sigma}^\prime \textbf{V}^{\prime H} \textbf{M}^{-1}
\textbf{V}^\prime)$ is a monotonically increasing function of $y$.
The monotonicity of functions $\text{tr} (\mathbf{\Sigma}^\prime \textbf{V}^{\prime H} \textbf{M}_\text{x}
\textbf{V}^\prime)$ and $\text{tr} (\mathbf{\Sigma}^\prime \textbf{V}^{\prime H} \textbf{M}^{-1}
\textbf{V}^\prime)$ will not change if $\mathbf{\Sigma}^\prime$ and $\textbf{V}^\prime$
are replaced by $\mathbf{\Sigma}$ and ${\textbf{V}}_y$, respectively. This completes the proof.

\end{document}